\documentstyle[multicol,aps,prb,epsf]{revtex}

\begin{document}

\title{Application of a continuous time cluster algorithm  to the
  Two-dimensional\\
Random Quantum Ising Ferromagnet}

\author{Heiko Rieger}
\address{HLRZ, Forschungszentrum J\"ulich, 52425 J\"ulich, Germany}

\author{Naoki Kawashima} 
\address{Department of Physics, Toho University, Miyama 2-2-1, 
Funabashi 274, Japan}

\date{November 11, 1997}

\maketitle

\begin{abstract}
A cluster algorithm formulated in continuous (imaginary) time
is presented for Ising models in a transverse field. It works directly
with an infinite number of time-slices in the imaginary time
direction, avoiding the necessity to take this limit explicitly. The
algorithm is tested at the zero-temperature critical point of the pure
two-dimensional (2$d$) transverse Ising model. Then it is applied to
the 2$d$ Ising ferromagnet with random bonds and transverse fields,
for which the phase diagram is determined. Finite size scaling at the
quantum critical point as well as the study of the quantum
Griffiths-McCoy phase indicate that the dynamical critical exponent is
infinite as in 1$d$.
\end{abstract}

%\vskip0.5cm


\newcommand{\bc}{\begin{center}}
\newcommand{\ec}{\end{center}}
\newcommand{\be}{\begin{equation}}
\newcommand{\ee}{\end{equation}}
\newcommand{\beqn}{\begin{eqnarray}}
\newcommand{\eeqn}{\end{eqnarray}}

\begin{multicols}{2}
\narrowtext
\parskip=0cm

Quantum phase transitions in random transverse Ising models at and
close to their quantum critical point at zero temperature ($T=0$) have
attracted a lot of interest recently \cite{review}. In particular in
one dimension many astonishing results have been obtained with
powerful analytical (real space renormalization group \cite{fisher},
field theory \cite{mckenzie}) and numerical (exact diagonalization
after mapping on a free fermion model \cite{youngrieger,igloirieger})
tools. Among the most important results is that {\it at} the critical
point time scales diverge exponentially fast, implying that the
dynamical exponent is $z_{\rm crit}=\infty$.  Such a behavior is
reminiscent of thermally activated dynamics in classical random field
systems \cite{rfim}, and has also been proposed to be present at the
Anderson-Mott transition of disordered electronic systems at the
quantum phase transition \cite{kirk}.

The other important result for random transverse Ising models in one
dimension is that close to the quantum critical point there is a whole
region in which various susceptibilities diverge for $T\to0$ and that
all these singularities, also called Griffiths-McCoy singularities
\cite{griffiths,mccoy}, can be parameterized by a single dynamical
exponent $z(\delta)$ that varies continuously with the difference
$\delta$ to the critical pont and {\it diverges} for $\delta\to0$. It
is not clear in how far these properties, i.e.\ $z_{\rm crit}=\infty$
and $z(\delta\to0)=z_{\rm crit}$ also apply in higher dimensions: In 2
and 3-dimensional transverse Ising {\it spin glass} models a {\it
finite} value for $z_{\rm crit}$ has been reported \cite{qsg} and for
finite dimensional bond {\it diluted} ferromagnets it has been shown
\cite{senthil} that $z_{\rm crit}$ is infinite only at the percolation
threshold.

In this letter we therefore consider the two-dimensional random
ferromagnet (without dilution) in a transverse field with the help of
a new Monte-Carlo cluster algorithm that is particularly suited to
handle the inherent difficulties in the study of such a random quantum
system: The origin of the Griffiths-McCoy singularities are strongly
coupled clusters (with strong ferromagnetic bonds and weak transverse
fields) which are extremely hard to equilibrate in a conventional
Monte-Carlo algorithm, therefore we had to use a cluster method.
Moreover, the exponent $z(\delta)$ is a {\it non-universal} quantity
for which reason we really have to perform the so-called Trotter-limit
explained below.  In a related work, Pich and Young \cite{pich} have
studied essentially the same model with a discrete time cluster
algorithm, {\it not} performing the Trotter limit and therefore
concentrating on the {\it critical} behavior.

A continuous time algorithm that incorporates this limit right from
the beginning (in the spirit of ref.\ \cite{bw,worms}) is the most
efficient method we can think of. In the first part of this letter we
present the method we use and apply it to the pure case in two
dimensions, in the second part we present our results for the random
case. Details and additional results will be published elsewhere
\cite{kawashima}.

The system we are interested in is defined by the quantum mechanical
Hamiltonian
\be
{\cal H}
=-\sum_{\langle ij\rangle} J_{ij}\sigma_i^z\sigma_j^z
 +\sum_i\Gamma_i\sigma_i^x
\label{ham}
\ee
in which $\sigma_i$ are spin-$\frac12$ operators located on {\it any}
$d$-dimensional lattice (belw we consider a $L\times L$ square lattice
with periodic boundary conditions), $\langle ij\rangle$ indicate all
nearest neighbor pairs on this particular lattice, $J_{ij}\ge0$ {\it
  ferromagnetic} interactions (either uniform $J_{ij}=J$ or random)
and $\Gamma_i$ transverse fields (either uniform $\Gamma_i=\Gamma$ or
random -- note that the sign of $\Gamma_i$ can always be gauged away
by a local spin rotation).

To derive our continuous time algorithm we use the Su\-zu\-ki-Trotter
decomposition to represent the free energy ${\cal F}$ of the system
(\ref{ham}) at inverse temperature $\beta=1/T$ as the {\it limit} of a
($d$+$1$)-dimensional classical Ising model \cite{suzuki}:
\beqn
{\cal F} & = & -\beta^{-1}\lim_{\Delta\tau\to0}
\ln\,{\rm Tr}\exp(-{\cal S_{\rm class}})\;, \label{class} \\
{\cal S_{\rm class}} & = &
-\sum_{\tau,\langle ij\rangle} K_{ij} S_{i}(\tau) S_{j}(\tau)
-\sum_{\tau, i} K'_i S_{i}(\tau) S_{i}(\tau+1)\;.\nonumber 
\eeqn
Here the additional index $\tau=1,\ldots,L_\tau$ of the now {\it
classical} Ising spin variables $S_i(\tau)=\pm1$ labels the $L_\tau$
$d$-dimensional (imaginary) time slices within which spins interact
via $K_{ij}=\Delta\tau J_{ij}$ and among which they interact with
strength $K_i'=-\frac12\ln\tanh\Delta\tau\Gamma_i$. The number of time
slices $L_\tau$ is related to $\Delta\tau$ by
$\Delta\tau=\beta/L_\tau$, so that the limit $\Delta\tau\to0$ in
(\ref{class}) implies $L_\tau\to\infty$. 

%This means that couplings in
%the imaginary time direction get infinitely strong in this limit
%whereas couplings in the space direction get infinitely weak. Due to
%the former property the equilibration becomes extremely hard in a
%single spin flip algorithm for small values of $\Delta\tau$.

Taking the limit $\Delta\tau\to0$ consecutive spins with the same
value along the imaginary time direction, e.g.\
$S_i(\tau)$=$S_i(\tau+1)$=$\cdots$=$S_i(\tau+N)$, form continuous
{\it segments} $\overline{S}_i\{[\tau,\tau+t]\}$ of length
$t=N\cdot\Delta\tau$ rather than individual lattice points. Since we
are going to take this limit implicitly we will have to consider these
imaginary time segments as the dynamical objects in a Monte-Carlo
algorithm, and not the individual spin values at discrete imaginary
times any more. These segments correspond to continuous, uninterrupted
pieces of a spin's world line during which it has the same value, 
say $+1$, and its two ends, in the following called {\it cuts},
are those times when this spin changes to another value, say
$-1$.

As the next step we apply the scheme of the Swendsen-Wang cluster
update method \cite{swendsen} within the aforementioned implicit
continuous time limit. Remember that in this method in order to
construct the clusters to be flipped at random, neighboring spins
pointing in the same direction are connected with a certain probability
$p$: for neighbors in the space direction, for instance $S_i(\tau)$
and $S_j(\tau)$ with $\langle ij\rangle$ being nearest neighbors, it
is $p_{ij}=1-\exp(-2K_{ij})=2\Delta\tau J_{ij}+{\cal O}(\Delta\tau^2)$
and for neighbors in imaginary time, for instance $S_i(\tau)$ and
$S_i(\tau+1)$, it is $p'_i=1-\exp(-2K'_i)= 1-\Delta\tau\Gamma_i+{\cal
O}(\Delta\tau^2)$. These spin connection probabilities are now
translated into probabilities for creating (cutting) and connecting
segments.

\begin{figure}
\epsfxsize=\columnwidth\epsfbox{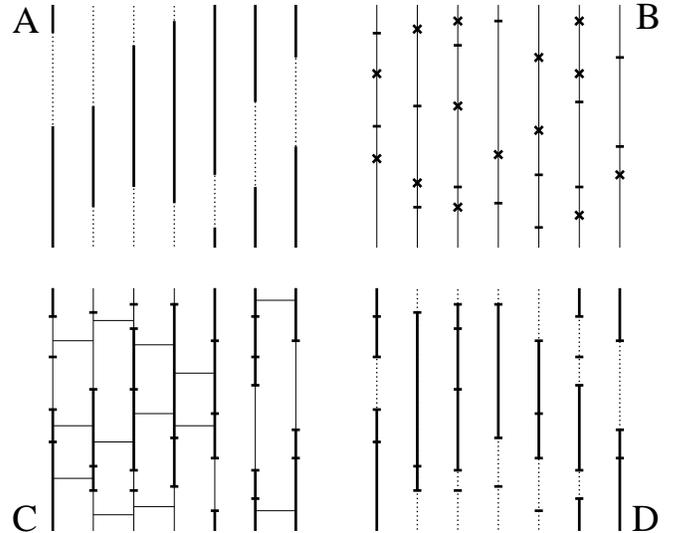}
\caption{
\label{cluster}
Sketch of the cluster construction procedure: a) Configuration of
segments before update --- full (broken) lines correspond to worldline
segments with spin up (down).  b) Insertion of new cuts (crosses)
in addition to old ones (bars) according to a Poisson process along each
worldline described by (\protect{\ref{probi}}).  c) Connection of
segments with probabilities given by (\protect{\ref{probij}}). d)
Configuration of segments after assigning randomly a spin value to
each cluster of connected segments. Before starting a new cycle (with A)
the redundant cuts {\it within} segments are removed.
}
\end{figure}

The probability for connecting spins along the imaginary time
direction at a particular site $i$ over a finite time interval of
length $t<\beta$ is given by the probability to set $t/\Delta\tau$
bonds, i.e.\ in the limit $\Delta\tau\to0$
\be
p_i'^{t/\Delta\tau}
=(1-\Delta\tau\Gamma_i)^{t/\Delta\tau}\to\exp(-\Gamma_i t)\;.
\label{probi}
\ee
This means for each site one generates new cuts in addition to the old
ones from the already existing segments via a Poisson process with
decay time $1/\Gamma_i$ along the imaginary time direction. Next one
connects segments on neighboring sites $i$ and $j$ that have the same
state and a nonvanishing time-overlap $t$, e.g.\ 
$\overline{S}_i\{[t_1,t_2]\}$ and $\overline{S}_j\{[t_3,t_4]\}$ with
$t$ now the length of the interval $[t_1,t_2]\cap[t_3,t_4]$). The
probability for {\it not} connecting the two segments is given by
\be
(1-p_{ij})^{t/\Delta\tau}
=(1-2\Delta\tau J_{ij})^{t/\Delta\tau}\to\exp(-2J_{ij}t)
\label{probij}
\ee
in the limit $\Delta\tau\to0$. Finally, one identifies clusters of
connected segments and assigns each of them (and herewith all of the
segments belonging to it) one value $+1$ or $-1$ with equal
probability. In Fig. \ref{cluster} the individual steps of this
cluster update procedure are illustrated. The measurement of
observables is straightforward: for instance the local magnetization
$m_i$ at site $i$ for one particular configuration of segments is
simply given by the difference between the total length of all
$+$segments and the total length of all $-$segments, divided by
$\beta$. The expectation value for the local susceptibility is
\be
\chi_i(\omega=0)
=\int_0^\beta d\tau\,\langle\sigma_i^z(\tau)\sigma_i^z(0)\rangle_{\rm QM}
=\beta\langle m_i^2\rangle_{\rm MC}\;,
\ee
where $\langle\cdots\rangle_{\rm QM}$ is the quantum mechanical
expectation value for model (\ref{ham}) and $\langle\cdots\rangle_{\rm
MC}$ is an average over all configurations generated during the
Monte-Carlo run.

For an actual implementation of the algorithm one has to provide
sufficient memory in which the information about the segments
$\overline{S}_i\{[t_1,t_2]\}$ (site index $i$, starting point $t_1$,
end point $t_2$ and state $+1$ or $-1$) is stored in linked
lists. The number of segments, which of course fluctuates, is
typically of the order ${\cal O}(\beta\Gamma_{\rm max}L^d)$, whereas
in discrete time algorithm the number of spin values to be
stored is $\beta L^d/\Delta\tau$, which diverges in the limit
$\Delta\tau\to0$. 

We tested our code first for the one-dimensional pure case ($d=1$,
$J_{ij}=1$ and $\Gamma_i=\gamma$), for which the critical
properties at $T=0$ are identical to those of the classical
2-dimensional Ising model and are well known exactly \cite{twod}
($\Gamma_c=1$, $\nu=1$, $\beta=1/8$, $z=1$).

As a first non-trivial application we considered the {\it pure} case
(again $J_{ij}=1$ and $\Gamma_i=\Gamma$) in {\it two} dimensions,
where the quantum critical point at $\Gamma_c$ is in the universality
class of the classical 3-dimensional Ising model.  Close to this
quantum critical point quantities are expected to obey finite size
scaling forms
\be
\langle{\cal O}\rangle
=L^{x_{\cal O}}\tilde{q}(L^{1/\nu}\delta,L^z/\beta)
\label{fss}
\ee
with $\delta=\Gamma-\Gamma_c$ and $x_{\cal O}$ the finite size scaling
exponent of the quantity ${\cal O}$. From the equivalence with the
3-dimensional classical Ising model one knows the dynamical exponent
{\it a priori} to be $z=1$, thus we can perform conventional
one-parameter finite size scaling if one chooses the aspect ratio
$\beta/L$ to be constant. In Tab.\ I we list the results of the finite
size scaling analysis of our data for system sizes up to $L=32$ (and
$\beta=8$). Our estimates for the critical field value
$\Gamma_c=3.044(1)$ and the thermal exponent $\nu=0.625(5)$ agree well
with the series expansion result \cite{series} and a recent DMRG study
\cite{dmrg}.

\end{multicols}
\renewcommand{\theequation}{\Alph{section}\arabic{equation}}
\widetext
%\noindent\rule{20.5pc}{.1mm}\rule{.1mm}{2mm}\hfill
\begin{table}
\begin{tabular}{ c || l || c | c | c | c  } 
      &  FSS form & $\Gamma_c$  &   $\nu$  & $\beta/\nu$    &
      $\gamma/\nu$   \\
\hline
m     &  $L^{-\beta/\nu}\tilde{m}(L^{1/\nu}(\Gamma-\Gamma_c))$
                 &  3.0440(2)  & 0.622(3) & 0.5050(5)      & [ 1.990(1) ] \\
$\chi$&  $L^{\gamma/\nu}\tilde{\chi}(L^{1/\nu}(\Gamma-\Gamma_c))$
                 &  3.0437(1)  & 0.621(1) & [ 0.500(3) ]   & 2.000(5)     \\ 
g     &  $\tilde{g}(L^{1/\nu}(\Gamma-\Gamma_c))$
                 & 3.0435(2)   & 0.629(5) &   ---          &   --- \\
\end{tabular}
\caption{
\label{puretable}
Estimates for the critical field strength and the critical exponents
for the quantum phase transition of the {\it pure} two-dimensional transverse Ising
model. The left column indicates the quantity for which the finite
size scaling analysis with constant aspect ratio $\beta=L/4$ has been 
performed ($m$ = magnetization, $\chi$ = uniform susceptibility and
$g$ = dimensionless ratio of moments). The values in brackets are
obtained from the scaling relation $\gamma/\nu=d+z-2\beta/\nu$ ($d=2$
and $z=1$ here).}
\end{table}
%\hfill\rule[-2mm]{.1mm}{2mm}\rule{20.5pc}{.1mm}
\begin{multicols}{2} 
\narrowtext
\noindent 

The real challenge for our algorithm (and our main motivation for
implementing it) is the {\it random} case, which we consider now. The
ferromagnetic couplings as well as the transverse field are now
quenched random variables, which we define both as being uniformly
distributed, i.e.\ 
\beqn
P(J_{ij})&=&\cases{1,&for $0<J_{ij}<1$\cr
                0,&otherwise\cr}\;\nonumber\\
P'(\Gamma_i)&=&\cases{\Gamma^{-1},&for $0<\Gamma_i<\Gamma$\cr
                0,&otherwise\cr}\;.
\label{uniform}
\eeqn
Observables now have to be averaged over a large number
($10^3$--$10^4$) of disorder configurations, which will be indicated
by $[\ldots]_{\rm av}$.

\begin{figure}
\epsfxsize=\columnwidth\epsfbox{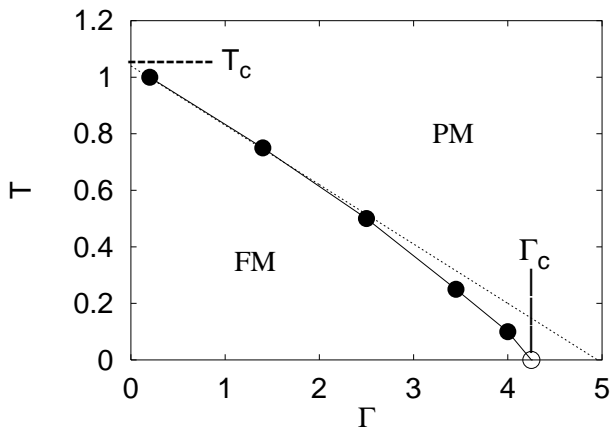}
\caption{
\label{phase}
The phase diagram of the two-dimensional {\it random} transverse Ising
model.  PM means paramagnetic, FM means ferromagnetic, $T_c=1.05(1)$ is the
critical temperature of the classical random Ising ferromagnet with a
uniform bond distribution, and $\Gamma_c=4.2(2)$ the location of the quantum
critical point we are interested in.}
\end{figure}

The main interest in our investigation is the determination of the
dynamical exponent at and above the quantum critical point. 
Since both are unknown, it is hard to work with a finite size
scaling form like (\ref{fss}). 
%\cite{remark}
Therefore we have chosen
the following procedure: first we determine the ($T$,$\Gamma$) phase
diagram of the model (\ref{ham}) with (\ref{uniform}) by calculating
the averaged ratio of moments $[g]_{\rm av}$ for various system sizes $L$ at
fixed temperature $T>0$ and $\Gamma$. At any finite temperature the
system is expected to be in the universality class of the classical
two-dimensional classical random bond Ising ferromagnet. For small
temperatures the crossover region to the quantum universality class
becomes larger, necessitating large system sizes (we went up to
$L=32$) to get a reliable estimate of $\Gamma_c(T)$. By extrapolating
the latter to $T=0$, see Fig. \ref{phase}, we obtain for the location of the
quantum critical point $\Gamma_c=\Gamma_c(T=0)=4.2\pm0.2$.

For $\Gamma=\Gamma_c$ it is $\delta=0$ and any observable is expected
to be a function of the aspect ratio $\beta/L^z$ alone. On the other
hand, {\it if} the two-dimensional case, which we consider here, is
similar to the one-dimensional case where one has unconventional (or
activated) scaling \cite{fisher,mckenzie,youngrieger,igloirieger} one
would expect $z=\infty$, which implies that $\ln\beta/L^{\psi}$ would
be the appropriate scaling variable (with $\psi$ a fit-parameter,
playing the role of the barrier exponent in classical activated
dynamics \cite{rfim,kirk,senthil2}). Our data scale much better in
according to the latter scenario with $\psi$ close to $0.5$ as in 1d
(details will be published in \cite{kawashima}), in accordance with the
results reported by Pich and Young \cite{pich}.

Now we turn our attention to the Griffiths-McCoy region in the
disordered phase ($\Gamma>\Gamma_c$). Due to the presence of
strongly coupled regions in the system the probability distribution of
excitation energies (essentially inverse tunneling times for these
ferromagnetically ordered clusters) becomes extremely broad. As a
consequence we expect the probability distribution of local
susceptibilities to have an algebraic tail at $T=0$
\cite{youngrieger,qsg_griff}
\be
\Omega(\ln\chi_{\rm local})
\approx-\frac{d}{z(\Gamma)}\ln\chi_{\rm local}
\label{localsus}
\ee
where $\Omega(\ln\chi_{\rm local})$ is the probability for the
logarithm of the local susceptibility $\chi_i$ at site $i$ to be
larger than $\ln\chi_{\rm local}$. The dynamical exponent $z(\Gamma)$
varies continuously with the distance from the critical point and
parameterizes the strengths of the Griffiths-McCoy singularities also
present in other observables. At finite temperatures $\Omega$ is
chopped off at $\beta$, and close to the critical point one expects
finite size corrections as long as $L$ or $\beta$ are smaller than the
spatial correlation length or imaginary correlation time,
respectively. We used $\beta\le1000$ and averaged over at least 512
samples.

\begin{figure}
\epsfxsize=\columnwidth\epsfbox{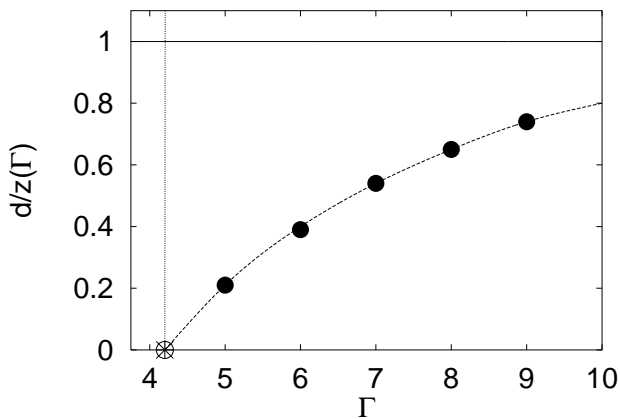}
\caption{
\label{dynexp}
The value of $d/z(\Gamma)$ obtained from analyzing the integrated
probability distribution of $\ln\chi_{\rm local}$ according to
(\protect{\ref{localsus}}) in the Griffiths-McCoy region. The vertical
line indicates the (approximate) region of the critical point at
$\Gamma_c\sim4.2$, the open circle corresponds to $z(\Gamma_c)=\infty$
and the horizontal line at $d/z=1$ indicates the expected limit
$\lim_{\Gamma\to\infty}z(\Gamma)=d$. The broken line is just a guide
to the eye.}
\end{figure}

In Fig. \ref{dynexp} we show our result for $d/z(\Gamma)$ in the
Griffiths-McCoy region. For $\Gamma\to\infty$ we expect
$d/z(\Gamma)\to1$, since this is the results for {\it isolated} spins
in random fields with non-vanishing probability weight at
$\Gamma_i=0$. The more interesting limit is $\Gamma\to\Gamma_c$. Due
to the aforementioned finite size effects an accurate estimate of
$d/z$ becomes increasingly hard approaching the critical point.
Nevertheless, the data are well compatible with
$\lim_{\Gamma\to\Gamma_c} z(\Gamma)=\infty$, implying that here, in
analogy to the one-dimensional case \cite{fisher,youngrieger} this
limit and the {\it critical} dynamical exponent $z_{\rm crit}$ agree.

To summarize we have presented a new Monte-Carlo cluster algorithm in
continuous imaginary time with which we studied the random transverse
Ising model in two dimensions. We determined the
temperature-transverse field phase diagram, estimated the location of
the quantum critical point at zero temperature and performed a finite
size scaling analysis at the critical point. Here we found indications
for an exponential divergence of time scales ($z=\infty$) and also the
dynamical exponent parameterizing the strength of the quantum
Griffiths-McCoy singularities extrapolates to
$z(\Gamma\to\Gamma_c)=\infty$. In this respect the phenomenology of
the one-dimensional model seems to extend to higher dimension. Another
aspect is the different scaling behavior of average and typical
correlations in the one-dimensional case, which seems to be present in
2d \cite{pich}, too.

We thank F.\ Igl\'oi and A.\ P.\ Young for helpful discussions. H.\ 
R.'s work was supported by the Deutsche Forschungsgemeinschaft (DFG)
and is grateful to the Toho University Department of Physics for kind
hospitality.  N.K.'s work was supported by the grant-in-aid
(No.09740320) from the ministry of education, science and culture.

\end{multicols}
\end{document}